\documentclass[global,twocolumn]{svjour}

\usepackage{latexsym}
\usepackage{graphics}
\usepackage{color}
\usepackage[draft]{hyperref}
\usepackage{amsmath}
\usepackage{graphicx}
\usepackage{color}
\usepackage{cuted}

\newcommand{\comment}[1]{}
\newcommand{\corrected}[1]{}

\journalname{Applied Physics B}

\begin{document}

\title{I.C.E.: a Transportable Atomic Inertial Sensor for Test in Microgravity}

\author{
    R. A. Nyman\inst{1} \and
    G. Varoquaux\inst{1} \and
    F. Lienhart\inst{2} \and
    D. Chambon\inst{3} \and
    S. Boussen\inst{2} \and
    J.-F. Cl\'{e}ment\inst{1} \and
    T. Muller\inst{4} \and
    G. Santarelli\inst{3} \and
    F. Pereira Dos Santos\inst{3} \and
    A. Clairon\inst{3} \and
    A. Bresson\inst{2} \and
    A. Landragin\inst{3} \and
    P. Bouyer\inst{1}
}

\institute{Laboratoire Charles Fabry de l'Institut d'Optique, Centre National de la Recherche Scientifique et Universit\'{e} Paris Sud 11, B\^{a}t. 503,
Campus Universitaire d'Orsay, 91403 Orsay Cedex, France
    \and
Office National d'Etude et de Recherches A\'{e}rospatiales, Chemin de la Huni\`{e}re, 91761 Palaiseau, France
    \and
LNE-SYRTE, UMR8630, Observatoire de Paris, 61 avenue de l'Observatoire, 75014 Paris, France
    \and
Institute for Quantum Optics, University of Hannover, Welfengarten 1, 30167 Hannover, Germany}
\date{Received: date / Revised version: date}
%
\maketitle

\begin{abstract}

We present our the construction of an atom interferometer for inertial sensing in microgravity, as part of the I.C.E. (\textit{Interf\'{e}rom\'{e}trie Coh\'{e}rente pour l'Espace}) collaboration. On-board laser systems have been developed based on fibre-optic components, which are insensitive to mechanical vibrations and acoustic noise, have sub-MHz linewidth, and remain frequency stabilised for weeks at a time. A compact, transportable vacuum system has been built, and used for laser cooling and magneto-optical trapping. We will use a mixture of quantum degenerate gases, bosonic $^{87}$Rb and fermionic $^{40}$K, in order to find the optimal conditions for precision and sensitivity of inertial measurements. Microgravity will be realised in parabolic flights lasting up to 20s in an Airbus. We show that  the factors limiting the sensitivity of a long-interrogation-time atomic inertial sensor are the phase noise in reference frequency generation for Raman-pulse atomic beam-splitters and acceleration fluctuations during free fall.

\end{abstract}

\section{Introduction}

Intense research effort has focussed on the study of degenerate quantum gases and macroscopic matter waves since their
first observation in 1995. Atom interferometers benefit from the use of trapped ultracold atomic gases, gaining  good
signal-to-noise ratios due to the high atomic densities, and the coherence required for the visibility of interference
patterns due to the low temperatures\cite{Bermann97}. The sensitivity of an interferometric measurement also depends on
the interrogation time, the time during which the sample freely evolves. This time is limited by both the free-fall of
the atomic cloud, requiring tall vacuum chambers, and by its free expansion, demanding extra-sensitive detection systems
for extremely dilute clouds. Ultralow temperatures further reduce the expansion. \corrected{From Franck: link between
sensitivity and expansion is not immediate - add an argument.}

In conceiving the next generation of extreme-precision atom interferometers, there is much to be gained by performing
experiments in microgravity \cite{Sleator99,Bongs04}. Free-fall heights of more than 100m, corresponding to durations of
5 seconds or more are available either in a drop tower (e.g. ZARM Bremen, Germany) or in a parabolic flight in an
aeroplane. Laboratory experiments are limited to about 300ms of free fall.\corrected{200ms changed to 300ms.} The
sensitivity of an interferometric accelerometer increases quadratically with time, and thus one can expect to gain more
than two orders of magnitude in having a transportable, drop-compatible device.

There remain questions over the best method to perform atom interferometry. Bosons suffer from interaction shifts
leading to systematic errors such as the clock shift, a problem not apparent in ultracold fermions\cite{Gupta03}.
However, degenerate fermions have an intrinsically broad momentum distribution due to Pauli blocking, limiting the
visibility of interference patterns. Furthermore, to achieve quantum degeneracy, fermions must be cooled using a buffer
gas, typically an ultracold gas of bosons, thus complicating experiments using fermions. Pairs of fermions (molecules or
Cooper pairs\cite{Regal04}) can be created by applying a homogeneous magnetic field (Feshbach
resonances\cite{LENS_Feshbach}), offering yet more possible candidate spe\-cies for atom interferometers.

A further bonus to free-fall is the possibility of using weaker confining forces for the atoms, since gravity need not
be compensated with additional levitation forces\cite{Leanhardt03}. Temperatures achieved by evaporative cooling and
adiabatic expansion are lowered as the trapping potential is reduced. Not only does the sensitivity of an
interferometric measurement benefit, but also new phases of matter may be observed if the kinetic energy can be made
smaller than the interatomic potential. A reduced-gravity environment will permit study of new physical phenomena, e.g.
spin dynamics and magnetic ordering (see for example \cite{Schmaljohann04} and references therein).

This article presents our design for a transportable, boson-fermion mixture, atom interferometer, compatible with a
parabolic flight in an aeroplane. We describe our laser systems: a temporary bench for ground-based development, and the
rack-mounted transportable system, based on frequency-doubled tele\-com\-muni\-cations lasers. We then explain our vacuum
system and optics for atomic manipulation, and the accompanying support structure. Finally we describe the
Raman-transition based atom-interferometric accelerometer, and show that the limits to in-flight performance are
vibrations (acceleration fluctuations) and phase-noise on the Raman laser frequency difference.

\subsection{Overview of the Experiment}

The central components of this project are the atomic-physics vacuum system, the optics, and their supports. The atomic manipulation starts with alkali-metal vapour dispensers for rubidium and potassium\cite{SAES_Getters}. A slow jet of
atoms is sent from the collection chamber by a dual-species, two-dimensional, magneto-optical trap (2D-MOT) to the
trapping chamber, for collection and cooling in a 3D-MOT. Atoms are then be transferred to a conservative, far-off-resonance optical-dipole trap (FORT) for further cooling towards degeneracy. The sample is then ready for
coherent manipulation in an atom-interferometer. Raman two-photon transition will be used as atomic beam-splitters and mirrors. Three-pulse sequences ($\pi/2 - \pi - \pi/2$) will be used for accelerometry.

All light for the experiment arrives by optical fibres, making the laser sources independent of the vacuum system. Transportable fibred laser sources for laser cooling and trapping have been fabricated with the required frequency stability. The techniques for mechanically-stable power distribution by free-space fibre couplers function according to specifications. The vacuum chamber is compatible with the constraints of microgravity in an Airbus parabolic flight. Such a flight permits total interrogation times up to 7s, giving a potential sensitivity of better than $10^{-9}\,\rm{m}\,\rm{s}^{-2}$ per shot, limited by phase noise on the frequency reference for the Raman transitions.

\corrected{From Robert: This section has been substantially changed after comments from other authors.}

\section{Laser Systems}

\subsection{Ground-based laser diodes for Potassium and Rubidium cooling}

Our test laser system is not intended to fly, but nonetheless represents several technical achievements, detailed in
Ref. \cite{Nyman06}. All of the lasers and optical amplifiers for trapping and cooling light are built around commercial
semiconductor elements (\textit{Eagleyard}) with home-made mounts and drive electronics. Semiconductor technology is one
of the candidates for atomic-physics lasers in micro-gravity experiments: the chips are small, lightweight and robust,
with low power consumption.

Extended-cavity grating-diode lasers (based on a design by Arnold et al.\cite{Arnold98}) are locked to atomic
transitions (the hyperfine structure of the D2 lines of $^{87}$Rb and $^{39}$K, as appropriate), frequency shifted by
acousto-optical modulators, injected into tapered amplifiers, then input to the optical fibres. We produce more than
200mW of useful light (out of the fibres) for trapping and cooling each species for both the 2D-MOT and the 3D-MOT.

One major difficulty was in making the master oscillator at 766.5nm (potassium D2 transition, wavelength in air).
Semiconductor lasers at 780nm (rubidium D2 line) have been available for some time\cite{Wieman91}, but are less easily
found at short wavelengths. We pulled a 780nm diode to 766.5nm using very weak feedback, by anti-reflection coating the
output face, and ensuring low reflectance from the grating (which was optimised for UV not visible light). Decreasing
the feedback increases the threshold current, which increases the number of carriers in the active region, increasing
the energy of the lasing transitions, thus giving gain at relatively short wavelengths. The tapered amplifiers we use
work equally well for the two wavelengths.

\subsection{Continuous-Wave Fibre-Laser Source at 780 nm for Rubidium Cooling}
\label{source}

An entirely pigtailed laser source is particularly appropriate in our case as it does not suffer from misalignments due to environmental vibrations. Moreover, tele\-com\-muni\-cations laser sources in the C-band (1530--1570 nm) have narrow linewidths ranging from less than 1MHz for laser diodes, down to a few kHz for Erbium doped fibre lasers. By second-harmonic generation (SHG) in a nonlinear crystal, these $1.56\mu$m sources can be converted to 780nm sources \cite{Mahal96,Thompson03,Dingjan06}. Such devices avoid having to use extended cavities as their linewidths are sufficiently narrow to satisfy the requirements of laser cooling.

\begin{figure}
\resizebox{\columnwidth}{!}{%
  \includegraphics{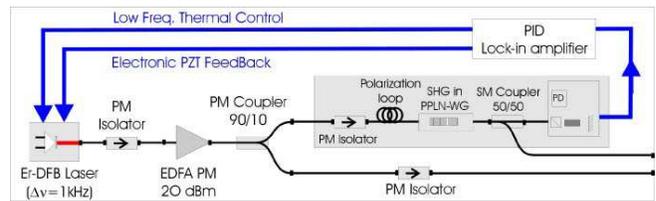}}
    \caption{Transportable laser set-up schematic. A double-loop feedback system is used for frequency control: the first returns a saturated absorption signal to the piezoelectric transducer; the second loop compensates thermal drifts of the fibre laser when the error signal of the first loop becomes large.}
    \label{fig:scheme}
\end{figure}

Our laser setup is sketched in Figure \ref{fig:scheme}. A 1560nm Erbium doped fibre laser is amplified by a 500mW
polari\-sation-main\-taining (PM) Erbium-doped fibre amplifier (EDFA). A 90/10 PM fibre-coupler directs $10\%$ of the pump power to a pigtailed output. $90\%$ of light is then sent into a periodically-poled Lithium-Niobate Waveguide (PPLN-WG). This crystal is pigtailed on both sides with 1560nm single-mode fibres. The input fibre is installed in a polarisation loop system in order to align the electric field with principal axes of the crystal. A fibre-coupler which is monomode at 780nm, filters pump light after the crystal and sends half of the 780nm light into a saturated- absorption spectroscopy device for frequency servo-control. The other half is the frequency-stabilised pigtailed output. The whole device, including the frequency control electronics was implemented in a rack for ease of transport. Typical output from the first generation device was 500$\mu$W of 780nm light, with more than 86dB attenuation of 1560nm light after 3m of monomode fibre. A more recent version ($>50$mW) has been used to power a magneto-optical trap. \corrected{Added: A more recent version ($>50$mW) has been used to power a magneto-optical trap.}

\begin{figure}
\resizebox{\columnwidth}{!}{%
  \includegraphics{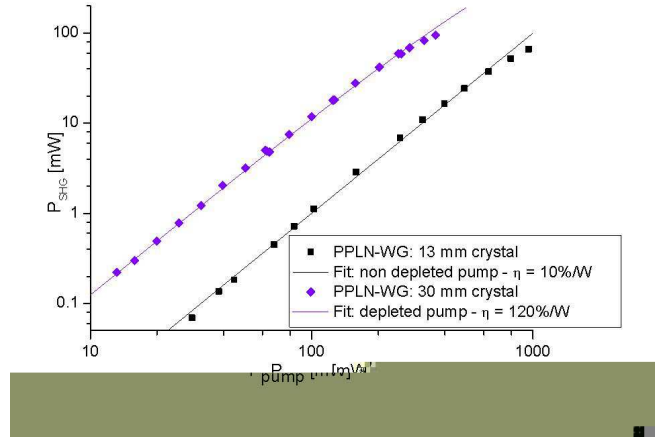}}
    \caption{Second-harmonic generation as a function of pump power. Two crystals (13mm, 30mm) were tested. Fits to non-depleted pump (13mm crystal, 10\%/W efficiency, squares) and depleted pump (30mm crystal, 120\%/W, lozenges).}
    \label{fig:PPLNWG}
\end{figure}

Two PPLN-WGs from HC-Photonics were tested. Both have a poling period appropriate for SHG at 780nm. They have the same quasi-phase matching temperature of $63^\circ$C. The first is 13mm long, doped with $1\%$ MgO, and is used in our
laser source. The second is 30mm long, doped with $5\%$ MgO. Figure \ref{fig:PPLNWG} gives the output power as a function of the pump power. The 13mm long crystal has a fibre-to-fibre efficiency of $10\%/$W. The fit curve corresponds to the non-depleted pump regime. Photorefractive effects appear around 10mW of 780nm light. In practice the laser is run with 100mW pump power. Power fluctuations in this crystal are due to two phenomenon: first the input fibre does not maintain polarisation, and polarisation fluctuations lead to a variation of the output power. Secondly the output fibre of the crystal is not single mode at 780nm. Thus the power distribution in the fundamental mode varies with time, leading to power fluctuations when the crystal is pigtailed to a single-mode fibre at 780nm. The second crystal has a fibre-to-fibre efficiency of $120\%/$W for low pump power. The fit curve corresponds to a
depleted regime. Photorefractive threshold is estimated around 60mW of second harmonic. The input fibre is still not
polarisation maintaining, leading to output power drifts, but the output fibre is PM and single mode at 780nm,
which greatly reduces power fluctuations.

\subsubsection{Frequency Stabilisation:}

\begin{figure}
\resizebox{\columnwidth}{!}{%
  \includegraphics{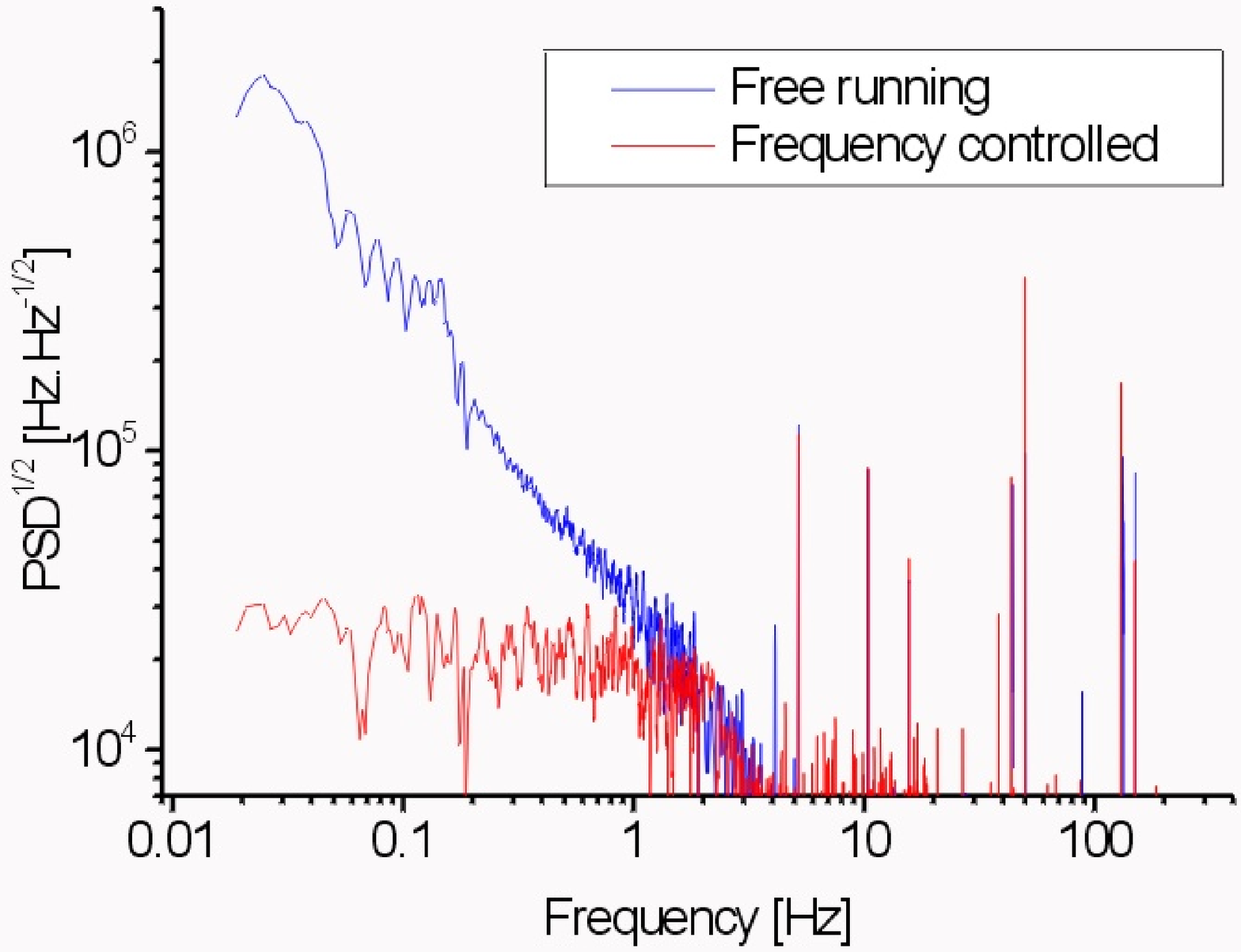}}
    \caption{Noise spectral density of the laser frequency in open and closed loop configurations. Data derived from error signal of frequency control system.}
    \label{fig:freqcontrol}
\end{figure}

A Doppler-free saturated-absorption spectroscopy system without polarisation sensitive elements provides the frequency reference signal. The frequency of the laser has been tested by locking to a crossover of $^{85}$Rb. The laser frequency is oscillated over a few 100kHz by modulating the piezoelectric element of the fibre Bragg grating of the pump laser. The modulation frequency is 1.3kHz, permitting long-term drifts to be compensated without significantly broadening the laser linewidth. \corrected{Removed sentence ``This modulation broadens the linewidth of the source'' but added word ''significantly``}. The spectroscopic signal is demodulated by a phase-sensitive detection and fed back to the piezo. Figure \ref{fig:freqcontrol} presents the spectral density of noise with and without frequency stabilisation. Noise up to 1.6Hz is attenuated, a frequency corresponding to the low-pass filter bandwidth of the demodulation. Points below $7\rm{kHz}/\sqrt{\rm{Hz}}$ are not represented because they are below the measurement noise. The r.m.s. frequency excursion in the band 0--20Hz is less than 200kHz. \corrected{Changed end of sentence from ``...band is 180kHz.''. Comment From Franck: Noise quoted doesn't quite add up with the figure.}

The laser remains frequency locked even with strong mechanical disturbances (hand claps, knocks on the rack ...), but cannot withstand even small variations of the ambient temperature. The fibre source at 1560nm, though temperature controlled, suffers frequency drifts due to temperature changes of the fibre. Small fluctuations are compensated by the frequency loop but long-term drifts are beyond the range of the piezoelectric servo-loop, so the laser jumps out of lock. An integrated circuit based on a PIC 16F84 micro-controller was developed: the output voltage of the regulator is monitored by the micro-controller, and, when fixed boundaries are exceeded, the set temperature of the laser controller is adjusted. This additional loop prevents the frequency control from unlocking without modifying the frequency properties of the source. The laser typically stays locked for up to three weeks.

\subsection{Fibre Power Splitters}

The optical bench and the vacuum chamber are not rigidly connected to each other, and laser light is transported to the vacuum chamber using optical fibres. Stability in trapping and coherent atom manipulation is assured by using only polarisation maintaining fibres. Six trapping and cooling laser beams are needed for the 3D-MOT and five for the 2D-MOT, with relative power stability better than a few percent. We have developed fibre beam-splitters based on polarising cubes and half-wave plates with one input fibre and the relevant number of output fibres. The stability of the beam splitters has been tested by measuring the ratio of output powers between different outputs as a function of time. Fluctuations are negligible on short time scales (less than $10^{-4}$ relative intensity over 1s), and very small over typical periods of experimental operation (less than 1\% over a day). Even over months, drifts in power distribution are only a few percent, which is sufficient for this experiment.

\section{Mechanical and Vacuum Systems}

\begin{figure}
\resizebox{\columnwidth}{!}{%
\includegraphics{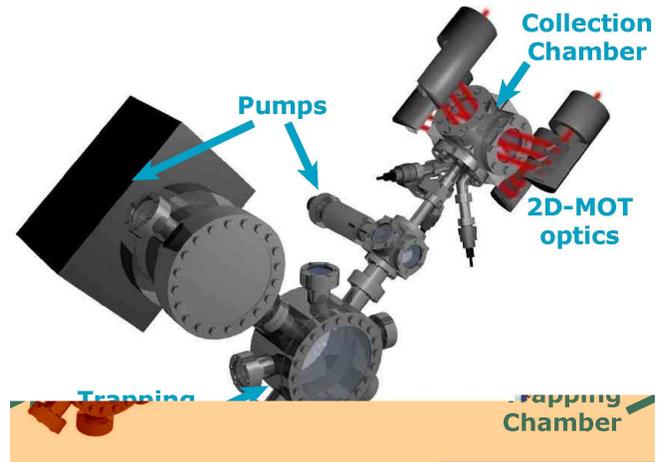}}
\caption{Artist's impression of the vacuum system. Atoms are transferred from the collection chamber, using a 2D-MOT,
to the trapping chamber, where they are collected in a 3D-MOT. The trapping chamber has large optical accesses for the
3D-MOT, optical-dipole trap (FORT), imaging, and interferometry. There is a getter pump between the two chambers to
ensure a large pressure difference. The other pump is a combined ion pump-titanium sublimation pump.}
\label{fig:chamber model}
\end{figure}

\begin{figure}
\resizebox{\columnwidth}{!}{%
\includegraphics{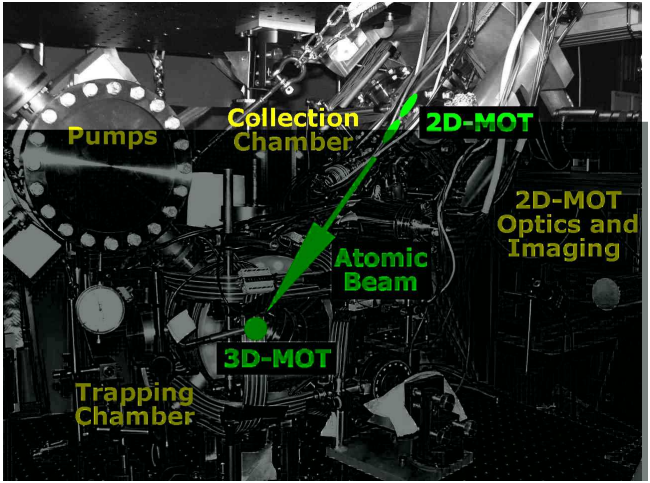}}
\caption{Photograph of the vacuum chamber, the support structure and the optics for magneto-optical traps.}
\label{fig:chamber photo}
\end{figure}

The mechanical construction of the apparatus is critical to any free-fall experiment. Atomic-physics experiments  require heavy vacuum systems and carefully aligned optics. Our design is based around a cuboidal frame of foam-damped hollow bars with one face being a vibration-damped optical breadboard: see Figs. \ref{fig:chamber model} and \ref{fig:chamber photo}. The outside dimensions are 1.2m $\times$ 0.9m $\times$ 0.9m, and the total weight of the final system is estimated to be 400kg (excluding power supplies, lasers, control electronics, air and water flow). The frame provides support for the vacuum system and optics, which are positioned independently of one another. The heavy parts of the vacuum system are rigged to the frame using steel chains and high-performance polymer slings under tension, adjusted using turnbuckles; most of the equipment being standard in recreational sailing or climbing. The hollow bars have precisely positioned grooves which permit optical elements to be rigidly fixed (bolted and glued) almost anywhere in the volume within the frame. An adaptation for transportability will be to enclose the frame in a box, including acoustic and magnetic shielding, temperature control, air overpressure (dust exclusion), as well as ensuring safety in the presence of the high-power lasers.

The vacuum chamber has three main parts: the collection chamber (for the 2D-MOT), the trapping chamber (for the
3D-MOT and the FORT) and the pumps (combined ion pump and titanium sublimation pump) Between the collection and trapping chambers there is an orifice and a getter pump, allowing for a high differential pressure, permitting rapid collection by the 2D-MOT but low trap losses in the 3D-MOT and FORT. The magnetic coils for the 2D-MOT are under vacuum, and consume just 5W of electrical power.

The main chamber has two very large viewports as well as seven side windows (and one entry for the atoms from the
2D-MOT). Thus there is plenty of optical access for the 3D-MOT, the FORT, imaging and interferometry. To preserve this
optical access, the magnetic coils are outside of the chamber, although this markedly increases their weight and power
consumption.

To avoid heating due to vibrations in the FORT optics, or measurement uncertainties due to vibrations of the imaging system, the trapping chamber is as close to the breadboard as possible. For laboratory tests, the breadboard is lowest, and the 2D-MOT arrives at 45$^\circ$ to the vertical, leaving the vertical axis available for addition of
interferometry for precise measurements, e.g a standing light wave. Around the main chamber, large electromagnet coils
in Helmholtz-configuration will be added, to produce homogeneous, stable fields up to 0.12T (1200G), or gradients up to
0.6T/m (60G/cm).

\subsection{2D-MOT}

The 2D-MOT is becoming a common source of cold-atoms in two-chamber atomic-physics experiments\cite{Dieckmann98}, and  is particularly efficient for mixtures \cite{Ospelkaus06} of $^{40}$K and $^{87}$Rb, if isotopically enriched dispensers are used. Briefly, a 2D-MOT has four sets of beams (two mutually orthogonal, counter-propagating pairs) transverse to the axis of the output jet of atoms, and a cylindrical-quadrupole magnetic field generated by elongated electromagnet pairs (one pair, or two orthogonal pairs). Atoms are cooled transverse to the axis, as well as collimated. Implicitly, only slow atoms spend enough time in the 2D-MOT to be collimated, so the output jet is longitudinally slow. The number of atoms in the jet can be increased by the addition of the push beam, running parallel to the jet: a 2D-MOT$^+$. Typically the output jet has a mean velocity below 30m\,s$^{-1}$, with up to 10$^{10}$ at.s$^{-1}$ of $^{87}$Rb and
10$^{8}$at.s$^{-1}$ of $^{40}$K.

Our design uses 40mW per species for each of the four transverse beams, each divided into two zones of about 20mm using non-polarising beam-splitter cubes, corresponding to about three times the saturation intensity for the trapping transitions. The push beam uses 10mW of power, and is about 6mm in diameter. Each beam comes from an individual
polarisation-maintaining optical fibre, with the light at 766.5nm and 780nm being superimposed on entry to the fibres.
The 2D-MOT is seen as two bright lines of fluorescence in the collection chamber.

\begin{figure}
\resizebox{\columnwidth}{!}{%
\includegraphics{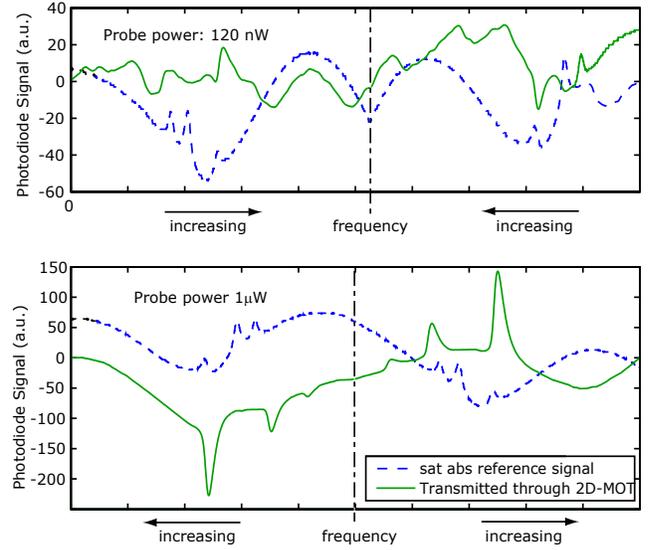}}
\caption{Absorption spectrum of atoms in the 2D-MOT in the collection chamber (green, solid line) and a reference
saturated-absorption spectroscopy signal (blue, dotted line). 120nW corresponds to about $I_{sat}/15$; $1\mu$W is
equivalent to $I_{sat}/2$. ($I_{sat}$ is the saturation intensity). Features with linewidth around 20MHz (equivalent to Doppler broadening of atoms moving at 20m.s$^{-1}$) are seen for the lowest probe powers, indicating a large density of slow atoms in the 2D-MOT. For higher probe powers, resonance light destroys the trapping in about 2ms, so the velocity
distribution is not resolved. Note that the inversion of absorption peaks is an artifact of the modulation-demodulation detection method.
}
\label{fig:2D-MOT spectrum}
\end{figure}

At the time of writing we do not have much quantitative data for the performance of our $^{87}$Rb 2D-MOT. One
interesting test we have performed is spectroscopy of the confined cloud, using a narrow probe beam parallel to the
desired output jet (replacing the push beam): see Fig \ref{fig:2D-MOT spectrum}. We detect a significant number of atoms in the 2D-MOT with velocities at or below 20m\,s$^{-1}$ (the output jet should have a similar velocity distribution). More sensitive spectroscopy is difficult, since the probe beam must be smaller than the transverse dimension of the atom cloud (less than 0.5mm) and much less than saturation intensity (1.6mW\,cm$^{-2}$), so as not to excessively perturb the atoms. We used a lock-in detection (modulation-demodulation-integration) method, averaging over many spectra. A saturated-absorption spectroscopy signal was used for calibration. We have not yet tested a $^{39}$K or $^{40}$K 2D-MOT.

\subsection{3D-MOT and Optical-Dipole Trap}

The atomic jet from the 2D-MOT is captured by the 3D-MOT in the trapping chamber. At the time of writing, we have
observed the transfer and capture of atoms, significantly increased by the addition of the push beam\cite{MOT_performance}. The 3D-MOT uses one polarisation-maintaining fibre input per species. Beams are superimposed and split into 6 arms (on a small optical breadboard fixed near one face of the frame) for the three, orthogonal, counter-propagating beam pairs. Once enough number of atoms are collected in the 3D-MOT, the 2D-MOT is to be turned off, and the 3D-MOT optimised for transfer to the FORT.

The FORT will consist of two, nearly-orthogonal ($70^\circ$) beams making a crossed, dipole trap using 50W of light at
1565nm. We will have rapid control over intensity using an electro-optical modulator, and beam size using a mechanical
zoom, after the design of Kinoshita et al.\cite{Kinoshita05}. Optimisation of transfer from the 3D-MOT to the FORT, and the subsequent evaporative cooling will require experiments. Strong, homogeneous, magnetic fields will be used to control interspecies interactions via Feshbach resonances\cite{LENS_Feshbach}, to expedite sympathetic cooling of $^{40}$K by $^{87}$Rb.

We can expect to load the 3D-MOT during less than 5s, then cool to degeneracy in the optical-dipole trap in around
3--10s. Thus we will be able to prepare a sample for interferometry in less than the free-fall time of a parabolic
flight (around 20s).

\begin{figure}
\resizebox{\columnwidth}{!}{%
\includegraphics{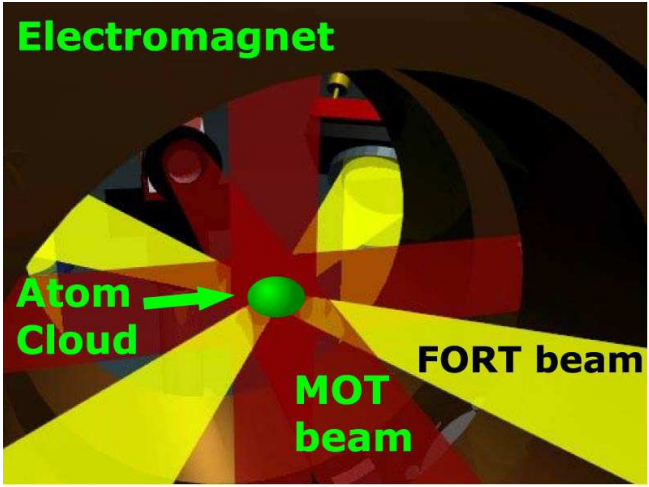}}
\caption{Artists impression of the 3D-MOT (dark, red beams, and the electromagnets) and Far-Off-Resonance Optical-Dipole
Trap (pale, yellow beams).
}
\label{fig:3D-MOT and FORT}
\end{figure}

\section{Performance}

\subsection{Coherent Raman-pulse Interferometer}

\corrected{This subsection title has changed. }

The acceleration measurement is based on an atomic interferometer using light pulses as beam splitters\cite{Chu91,Borde91}, a technique which has demonstrated best performance for atomic inertial sensors. Three Raman pulses ($\pi/2 - \pi - \pi/2$) to generate respectively the beam splitter, the mirror and the beam re-combiner of the atomic interferometer. Two counter-propagating lasers (Raman lasers) drive coherent transitions between the two hyperfine ground states of the alkaline atoms. Two partial wave-packets are created with differing momenta, due to absorption and stimulated emission of photons in the Raman lasers. The differences in momenta correspond to velocity differences of 1.2\,cm\,s$^{-1}$ for $^{87}$Rb and 2.6\,cm\,s$^{-1}$ for $^{40}$K for Raman lasers tuned close to the D2 lines. Finally, fluorescence detection gives a measurement of the transition probability from one hyperfine level to the other, given by $P=\frac{1}{2}(1-\cos(\Phi))$, where $\Phi$ being the interferometric phase difference. It can be shown\cite{Antoine03} that the interferometric phase difference depends only on the difference of phase between the Raman lasers at the classical position of the centre of the atomic wave-packets at the time of the pulses. In the case of an experiment in free fall, with no initial velocity of the atoms, the interferometric phase depends only on the average relative acceleration of the experimental apparatus with respect to the centre of mass of the free-falling atoms, taken along the direction of propagation of the Raman lasers. We neglect here the effects gradients of gravity on expanding and separating wave-packets, which cause small changes to the final fringe visibility. \corrected{We neglect here the effects gradients of gravity on expanding and separating wave-packets, which cause small changes to the final fringe visibility.}

As the measurement is performed in time domain with pulses of finite duration $\tau_R-2\tau_R-\tau_R$ separated by a free evolution time $T$, it is also sensitive to fluctuations of the relative phase of the Raman lasers between pulses. Moreover, as the measurement is not continuous but has dead time, the sensitivity of the interferometer is limited by an aliasing effect similar to the Dick effect in atomic clocks\cite{Dick87}. Thus, the sensitivity of the interferometer also depends on vibrations and on the phase noise on the beat note between the Raman lasers at multiples of the cycling frequency $T_c$. The effects of these noise sources is calculated\cite{Cheinet06} using the sensitivity function which gives the influence of the fluctuations of the Raman phase on the transition probability, and thus on the interferometric phase.

\subsection{Influence of Phase Noise}

\corrected{From Franck: Move first paragraph + figure to middle of third paragraph.}

The sensitivity of the interferometer can be characterised by the Allan variance of the interferometric phase  fluctuations, $\sigma^{2}(\tau)$, defined by:
\begin{eqnarray}
    \sigma_{\Phi}^{2}(\tau)&=&\frac{1}{2}\langle(\bar{\delta \Phi}_{k+1}-\bar{\delta \Phi}_{k})^{2}\rangle \nonumber\\
    &=&\frac{1}{2}\lim_{n\rightarrow \infty}\left\{
    \frac{1}{n}\sum_{k=1}^{n}(\bar{\delta \Phi}_{k+1}-\bar{\delta \Phi}_{k})^{2}\right\}
    \label{eq:variance_allan}
\end{eqnarray}
where $\delta \Phi$ is the fluctuation of the phase measured at the output of the interferometer, $\bar{\delta  \Phi}_{k}$ is the average value of $\delta \Phi$ over the interval from $t_{k}$ to $t_{k+1}$ (of duration $\tau$). For an interferometer operated sequentially at a rate $f_c=1/T_{\rm{c}}$, $\tau$ is a multiple of $T_c$, $\tau=m T_c$.

When evaluating the stability of the interferometric phase $\Phi$, one should take into account the fact that the measurement is pulsed. The sensitivity of the interferometer is limited only by the phase noise at multiples of the
cycling frequency weighted by the Fourier components of the transfer function. For large averaging times ($\tau \gg T_C$), the Allan variance of the interferometric phase is given by
\begin{equation}
\label{eq:dick} \sigma^{2}_{\Phi}(\tau)={\frac{1}{\tau}}\sum_{n=1}^{\infty}|H(2\pi n f_{\rm{c}})|^2
        S_{\phi}({2\pi n f_{\rm{c}}})
\end{equation}
where $S_{\phi}$ is the spectral power density of the phase difference between the Raman lasers.

Assuming square Raman pulses, the transfer function $H(f)$ of the Raman laser phase fluctuations to the interferometric phase is \cite{Cheinet06}:
\pagebreak
        
\begin{strip}
\begin{equation}
    \left|H(f) \right|^2 =
       \left|
            -\frac{4 \Omega \omega}{\omega^2-\Omega^2}\sin \left(\omega \frac{T+2\tau_R}{2}\right)
            \left(\sin\left(\omega \frac{T+2\tau_R}{2}\right) +
            \frac{\Omega}{\omega}\sin\left(\omega\frac{T}{2}\right)\right)
        \right|^{2}  \label{eq:Transfunct}
\end{equation}
\end{strip}
\hrulefill\par
\begin{figure}[h]
\resizebox{\columnwidth}{!}{%
\includegraphics{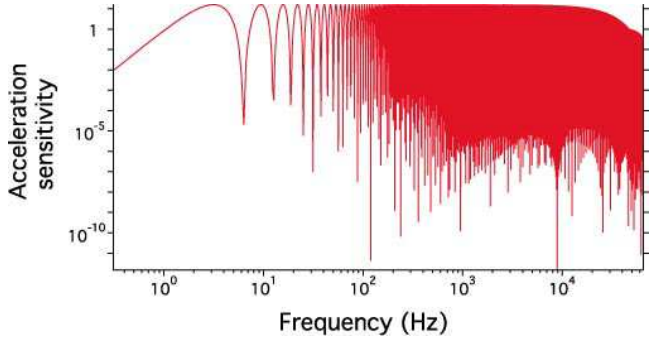}
}
\caption{Transfer function from amplitude of phase fluctuations to interferometric phase. The curve has been calculated for $T=0.5s$ between pulses, and pulse duration $\tau_R=50\mu$s. \comment{From Rob: What are the units of the vertical axis?}}
\label{transfer}
\end{figure}

\vskip 6 pt
\noindent
where $\omega = 2\pi f$ and  $\Omega$ is the Rabi oscillation frequency, taken in such way that the Raman $\pi$-pulses
have the ideal transfer efficiency: $\Omega = \pi/2\tau_R$. The transfer function is characterised by zeroes at multiples of 1/(T+2$\tau_R$) and decreases as 1/$\Omega^2$ for frequencies higher than the Rabi frequency, as illustrated in Figure \ref{transfer}.

For white phase noise $S_{\phi}^0$, and to first order in $\tau_R/T$, the phase stability is given by:
\begin{equation}
\label{whiteeq1} \sigma^{2}_{\Phi}(\tau)=\frac{\pi \Omega}{2}
S_{\phi}^0\frac{T_c}{\tau}.
\end{equation}
Thus the transfer function filters such noise for frequencies greater than the Rabi frequency: the shorter
the pulse duration $\tau_R$, and thus the greater the Rabi frequency, the greater the interferometer noise.  However, longer-duration pulses interact with fewer atoms (smaller velocity distributions) leading to an pulse duration, around $10\mu$s. More quantitatively, a desired standard deviation of interferometer phase below 1 mrad per shot, with pulse duration $\tau_R=10\mu$s, demands white phase noise of $4\times10^{-12}$ rad$^2$/Hz or less.

\corrected{Comment from Robert: Added sentence: ``However, longer-duration pulses interact with fewer atoms (smaller velocity distributions) leading to an pulse duration, around $10\mu$s.''}

\subsection{Generation of a Stable Microwave Source for Atom Interferometry}

\subsubsection{The 100 MHz Source Oscillator:}
The frequency difference between the Raman beams needs to be locked to a very
stable microwave oscillator, whose frequency is close to the
hyperfine transition frequency, $f_{MW}=6.834$ GHz for
$^{87}$Rb, and $1.286$ GHz for $^{40}$K. The reference frequency will be
delivered by a frequency chain, which transposes an RF source (typically a quartz oscillator) into the microwave domain, retaining the low level of phase noise. With degradation-free transposition the phase noise power spectral density of the RF oscillator, of frequency $f_{RF}$, is multiplied by $(f_{MW}/f_{RF})^2$.

\begin{figure}
\centerline{\resizebox{\columnwidth}{!}{%
  \includegraphics{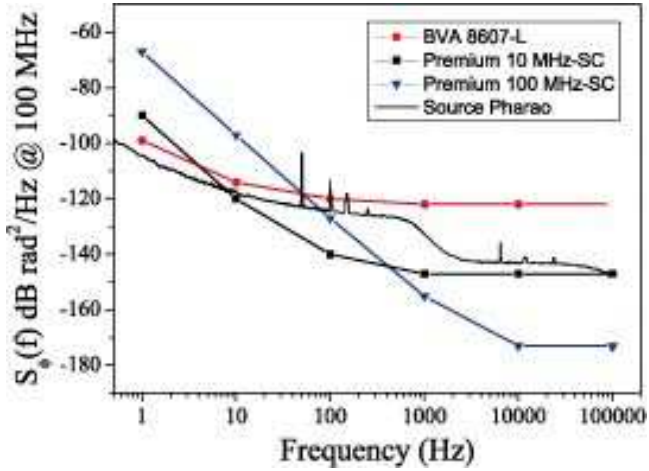}
}}
\caption{Specifications for the phase noise spectral power density of
different quartz oscillators, transposed at 100 MHz. The phase
noise of the source developed for the PHARAO project is also
displayed as a solid black line (courtesy of CNES).}
\label{fig:specquartz}
\end{figure}

No single quartz oscillator fulfills the requirements of very low
phase noise over a sufficiently large frequency range. We present
in figure \ref{fig:specquartz} the specifications of different
high stability quartz oscillators: a Premium 10 MHz-SC from Wenzel, a BVA
OCXO 8607-L from Oscilloquartz, and a Premium 100 MHz-SC quartz
from Wenzel. The phase noise spectral density is shown as transposed to 100 MHz, for fair comparison of the different oscillators.

The 100 MHz source we plan to develop for the ICE project will be
a combination of two phase-locked quartz oscillators: one at 100 MHz locked
onto one of the above-mentioned high-stability 10 MHz reference
oscillators. The bandwidth of the lock corresponds to the
frequency below which the phase noise of the reference oscillator is
lower than the noise of the 100 MHz oscillator.

The phase noise properties of such a combined source can be seen
in Figure \ref{fig:specquartz}, where we also show (solid line) the performance of the 100 MHz source
developed by THALES for the PHARAO space clock project. This
combined source has been optimized for mimimal phase noise at low
frequency, where it reaches a level of noise lower than any
commercially available quartz oscillator. An atomic clock is indeed mostly
limited by low-frequency noise, so
the requirements on the level of phase noise at higher frequency
($f>1$kHz) are less stringent than for an atom interferometer. A
medium performance 100 MHz oscillator is thus sufficient.

\corrected{From ONERA: Should mention that there is no noise added in going from 100MHz to 6.8GHz. NOTE from Rob: This point is discussed in the following subsubsection ``Frequency Chain''}

Using a simple model for the phase-lock loop, we calculated the phase noise spectral power density of the different combined sources we can make by locking the Premium 100 MHz-SC either to the Premium 10 MHz-SC (Source 1), or the BVA (source 2), or even the PHARAO source (source 3). We then estimated the impact on the interferometer of the phase noise of the 100 MHz source, assuming we are able to transpose the performance of the source at 6.8 GHz without degradation. The results presented in Table \ref{tb:noise} were calculated using Equation \ref{eq:dick} for the Allan standard deviation of the interferometric phase fluctuations for the different configurations and various interferometer parameters.
\begin{table*}
            \centering
            \begin{tabular}{|l|c|c|c|c|c|c|c|}
                    \hline

                                &            &Source 1&Source 2&Source 3&Best Source&Best Source\\
                    {\bf $T_c$} & {\bf $2T$}  & {\bf $\sigma_{\Phi}$($T_c$)} & {\bf $\sigma_{\Phi}$($T_c$)} & {\bf  $\sigma_{\Phi}$} & {\bf $\sigma_a$($T_c$)} & {\bf $\sigma_a$(1s) }\\
                    {\bf (s)} & {\bf (s)}  & {\bf (mrad)} & {\bf (mrad)} &  {\bf (mrad)} & {\bf (m.s$^{-2}$) / shot}&{\bf (m.s$^{-2}$.Hz$^{-1/2}$)}\\

                    \hline
                    \hline

                    0.25  & 0.1 &  1.2 & 3.5 & 2.2 &3x10$^{-8}$&1.5x10$^{-8}$\\
                    \hline
                    10  & 2  & 22 & 8.8 & 4.6 &1.1x10$^{-9}$&3.6x10$^{-9}$\\
                    \hline
                    10 & 5  & 55 & 20 & 10 &9.9x10$^{-11}$&3.1x10$^{-10}$\\
                     \hline
                    15 & 10 & 110 & 37 & 19 &4.7x10$^{-11}$&1.8x10$^{-10}$\\
                    \hline
             \end{tabular}
        \caption{Contribution of the 100 MHz source phase noise to the interferometric phase fluctuations
                    ($\sigma_\Phi$) and to the acceleration sensitivity ($\sigma_a$). The calculation has been
                    performed for a $^{87}$Rb interferometer, for each of the three different sources assuming pulse
                    duration  $\tau_R$=10 $\mu$s. $T_C$ is the cycle time for measurements, $2T$ is the total interrogation time. (Source 1: Premium; Source 2: BVA; Source 3: PHARAO)}
        \label{tb:noise}
        \end{table*}

For short interrogation times, such as $2T=100$ ms (the maximum interrogation time possible when the experiment is tested on the ground), Source 1 is best, whereas for long interrogation times, where the major contribution to the
noise comes from the lowest frequencies (0.1--10 Hz), Sources 2 and 3 are better.

We are currently using a source based on the design of Source 1 for the gravimeter experiment at SYRTE  \cite{Cheinet01}. Its performance is about 10\% better than predicted, as the reference oscillator phase noise level is lower than the specifications. Considering that the interferometer is intended for a zero-g environment, we plan to build a source based on Source 2.

We have assumed here that for any source, the phase noise below 1 Hz is accurately described as flicker noise, for which the spectral density scales as $S_{\phi}(f)= S_{\phi}(1\rm{Hz})/\it{f}^3$. If the phase noise behaves as pure flicker noise over the whole frequency spectrum, the Allan standard deviation of the interferometer phase scales as $T$. We note that the observer behaviour of the gravimeter is consistent with Table \ref{tb:noise}. \corrected{After comment from Franck: Change ``Such behaviour is observed in the gravimeter for long interrogation times''. to ``We note that the observer behaviour of the gravimeter is consistent with Table \ref{tb:noise}.''}

The sensitivity of the accelerometer improves with the square of the interrogation time, $T^2$. For example, for  $2T=10$s and $T_c=15$ s, the phase noise of Source 3 would limit the acceleration sensitivity of the interferometer to $4.6\times10^{-11}\rm{ms}^{-2}$ per shot for $^{87}$Rb. As the hyperfine splitting of $^{40}$K is five times smaller, the transposed phase noise is lower, and the measurement limit with $^{40}K$ decreases to $8.7\times10^{-12}\rm{ms}^{-2}$ per shot.

\subsubsection{The Frequency Chain:}
\begin{figure}[h]
\centerline{\resizebox{\columnwidth}{!}{%
  \includegraphics{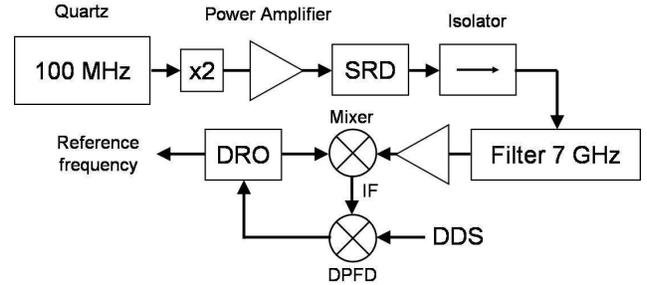}
}}
\caption{Scheme of the synthesis of the microwave reference signal. SRD: Step Recovery Diode. DDS: Direct Digital Synthesis. DRO: Dielectric Resonator Oscillator. DPFD: Digital Phase-Frequency Detector. IF: Intermediate Frequency.
}
\label{fig:setupchain}
\end{figure}

The microwave signal is generated by multiplication of the 100 MHz
source. We have developed a synthesis chain whose principle is
shown in figure \ref{fig:setupchain}.

The source is first frequency doubled, the 200 MHz output is filtered, amplified to 27 dBm, and sent to a Step Recovery Diode (SRD), which generates a comb of frequencies, at multiples of 200 MHz. An isolator is placed after the SRD in order to prevent back reflections to damage the SRD. The 35th harmonic (7 GHz) is then filtered (passed) using a passive filter. A dielectric resonator oscillator (DRO) is then phase locked onto the 7 GHz harmonic, with an adjustable offset frequency provided by direct digital synthesis. A tunable microwave source is thus generated which copies the phase noise of the 7 GHz tooth of the comb, within the bandwidth of the DRO phase-lock loop (about 500 kHz). The noise added by the frequency chain was measured by mixing the outputs of two identical chains, with a common 100  MHz source; this noise is weaker than the noise due to the 100 MHz source.

The derived contribution to the phase noise of a $^{87}$Rb interferometer is 0.6 mrad per shot for $\tau_R=10 \mu$s, $2T=10$s and $T_c=15$s. \corrected{Removed phrase ``, which is negligible with respect to the contribution from 100 MHz source.''} The sensitivity limit due to the frequency synthesis is almost negligible for the $^{40}K$. In conclusion, the limit to sensitivity comes predominantly from the phase noise of the low frequency oscillator. This contribution could be further reduced by the use of cryogenic sapphire oscillator \cite{Mann01}.\corrected{From Franck: Type : "In conclusion, ou finally, the limit in the sensitivity comes predominantly from the phase noise of the low freq osci. This contribution could be reduced by using a cryo sapphire osci."}

\subsection{Zero-Gravity Operation}
\begin{figure}[h]
\resizebox{\columnwidth}{!}{%
\includegraphics{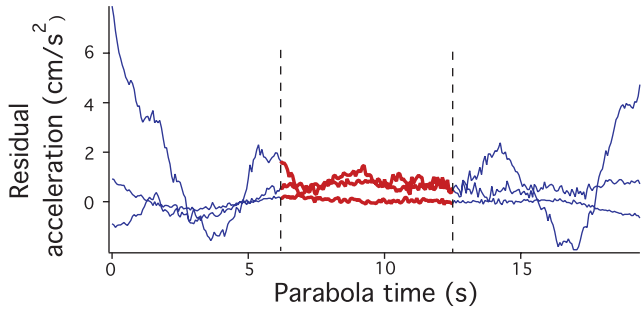}
}
\caption{Typical residual acceleration along three orthogonal axes during a parabolic flight. The period when an experiment can be performed in conditions of very low residual acceleration is highlighted. During this period (from 7 to 14 seconds, between then dotted lines) the apparatus may be allowed to float freely.}
\label{airacc}
\end{figure}

In this section, we estimate the possible limitations of the interferometer when used in a parabolic flight, by calculating the effect of residual acceleration in the Airbus (the proposed test vehicle for this experiment) during a parabola. During a typical flight the residual acceleration can be of the order of 0.1\,m\,s$^{-2}$, with fluctuations of acceleration of the same order (Fig. \ref{airacc}).

To determine the influence of environmental noise on the acceleration measurement, one uses the transfer function $H(f)$ for the phase (Equation \ref{eq:Transfunct}). Phase noise is equivalent to position noise, since the phase of the Raman beams is $\Delta\phi = k_L\delta z$, where $k_L$ is the wave-vector of the laser, $\Delta z$ the position difference along the laser path, and position is the second integral of acceleration over time. The variance of the fluctuation of the phase shift at the output of the interferometer is:
\begin{equation}
\sigma^2= \langle\left|\delta\left(\Delta\phi\right)\right|^2\rangle = k_L^2  \int_0^{\infty}S_a (f)\left|H(f)\right|^2/\omega^4 \,df
\label{var}
\end{equation}
where $S_a (f)$ is the acceleration-noise power density which corresponds to the Fourier transform of the temporal fluctuation. \corrected{From Franck: It seems that this equation is missing a factor $k_L^2$. From Rob: I've added the factor. I hope I've not made a mistake.}

\begin{figure}[h]
\resizebox{\columnwidth}{!}{%
\includegraphics[angle=0]{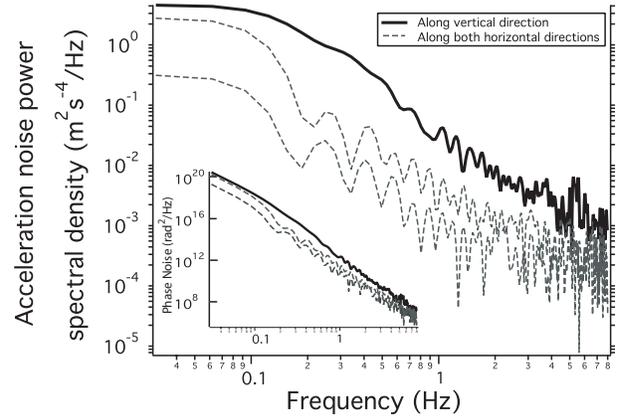}
}
\caption{Typical acceleration noise power spectral density during the quiet part of the zero-gravity parabola. The three curves represent the noise along three directions (vertical being the noisiest). Insert: the corresponding phase noise which should be taken into account in the actual interferometer performance.}
\label{spec2}
\end{figure}

From the residual acceleration curves for the Airbus, one can deduce the acceleration noise power in a bandwidth from 0.05 to 10Hz, giving an estimation of the noise on the measurement of acceleration. A spectral acceleration-noise power density curve for the useful low-noise part of a parabola is shown in Figure \ref{spec2}, and is converted to interferometric phase noise power spectral density by multiplication by $k_L^2 |H|^2/ \omega^4$. \corrected{From Franck: k2*H2/omega4. From Rob: I've added the factor. I still hope I've done it correctly.}

The vibration noise results in a substantial residual phase noise which is incompatible with the operation of
the accelerometer. Calculating the variance of the fluctuations from Equation \ref{var}, one obtains a variance $\sigma_{\Phi}\sim 10^7$\,rad, which corresponds to acceleration noise $\sigma_{a}\sim 1$\,m\,s$^{-2}$ where $\sigma_{\Phi}=k_L \sigma_{\rm a}T^2$ with $T=1$s. \corrected{From Rob: I have changed a symbol to $\sigma_a$ not $\sigma_\Phi$, and added a suitable value of $T$.} Thus a vibration isolation system will be required, reducing the noise by 60--80dB around 0.5 Hz, about 40dB at 50 Hz and less than 10dB beyond 1kHz. The situation can be more favourable if one restricts the measurements in the middle of the parabola, as indicated on the Figure \ref{airacc}.

\section{Conclusions}

We have shown our design for a transportable atom interferometer for parabolic flights in an Airbus. The device is built in two main parts, the laser systems and the atomic physics chamber. We have made major technical advances: high-stability frequency synthesis for coherent atom manipulation, flight-compatible laser sources and fibre power splitters, as well as a rugged atomic-physics chamber.

We have analysed the possibility of using this device in the micro-gravity environment of a parabolic flight, as a high-precision accelerometer, taking advantage of the long interrogation times available to increase the sensitivity to accelerations. We conclude that the limits to measurement under such conditions come from acceleration fluctuations and from phase noise in the frequency synthesis, and thus both aspects are to be minimised. Sensitivity of better than $10^{-9}\,\rm{m}\,\rm{s}^{-2}$ per shot is predicted. Comparisons of acceleration measurements made using two different atomic species (K and Rb) are possible.

\corrected{From Robert: The conclusion has been substantially changed after comments from other authors.}

The I.C.E. collaboration is funded by the CNES, as is RAN's fellowship. Further support comes from the European Union STREP consortium FINAQS.


\begin{thebibliography}{}
\bibitem{SAES_Getters} Natural-abundance dispensers are available from SAES Getters . We are interested in the isotopes $^{87}$Rb and $^{40}$K, which are respectively 28\% and 0.012\% naturally abundant. We will be investing in isotopically enriched $^{40}$K (about 5\%) dispensers in the near future.
\bibitem{MOT_performance} Our current best measurements are rather vague, suggest $>5 \times 10^7$ atoms in the cloud. However, the maximum size may in fact be several $10^9$ atoms. Loading to nearly the maximum atom number takes less than 5s.
\end{thebibliography}
\end{document}